\documentclass[aps,prl,twocolumn,superscriptaddress,showpacs,floatfix]{revtex4}
\usepackage{times}
\usepackage{graphicx}
\usepackage{epsfig}
\usepackage{amsmath}
\usepackage{amssymb}
\usepackage[version=3]{mhchem}
\usepackage{color}
\begin{document}

\title{Orbital Dimer Model for Spin-Glass State in Y{$_2$}Mo{$_2$}O{$_7$}}

\author{Peter M. M. Thygesen}
\affiliation{Department of Chemistry, University of Oxford, South Parks Road, Oxford OX1 3QR, U.K.}

\author{Joseph A. M. Paddison}
\email{paddison@gatech.edu}
\affiliation{Department of Chemistry, University of Oxford, South Parks Road, Oxford OX1 3QR, U.K.}
\affiliation{ISIS Facility, Rutherford Appleton Laboratory, Harwell Campus, Didcot, Oxfordshire OX11 0QX, U.K.}
\affiliation{School of Physics, Georgia Institute of Technology, 837 State Street, Atlanta, Georgia, 30332-0430, U.S.A.}

\author{Ronghuan Zhang}
\affiliation{Department of Chemistry, University of Oxford, South Parks Road, Oxford OX1 3QR, U.K.}

\author{Kevin A. Beyer}
\affiliation{Advanced Photon Source, Argonne National Laboratory, Argonne, Illinois 60439, U.S.A.}

\author{Karena W. Chapman}
\affiliation{Advanced Photon Source, Argonne National Laboratory, Argonne, Illinois 60439, U.S.A.}

\author{Helen Y. Playford}
\affiliation{ISIS Facility, Rutherford Appleton Laboratory, Harwell Campus, Didcot, Oxfordshire OX11 0QX, U.K.}

\author{Matthew G. Tucker}
\affiliation{ISIS Facility, Rutherford Appleton Laboratory, Harwell Campus, Didcot, Oxfordshire OX11 0QX, U.K.}
\affiliation{Diamond Light Source, Chilton, Oxfordshire, OX11 0DE, U.K.}
\affiliation{Spallation Neutron Source, Oak Ridge National Laboratory, Oak Ridge, Tennessee 37831, U.S.A.}

\author{David A. Keen}
\affiliation{ISIS Facility, Rutherford Appleton Laboratory, Harwell Campus, Didcot, Oxfordshire OX11 0QX, U.K.}

\author{Michael A. Hayward}
\affiliation{Department of Chemistry, University of Oxford, South Parks Road, Oxford OX1 3QR, U.K.}

\author{Andrew L. Goodwin}
\affiliation{Department of Chemistry, University of Oxford, South Parks Road, Oxford OX1 3QR, U.K.}

\date{\today}
\begin{abstract}
The formation of a spin glass usually requires both structural disorder and frustrated magnetic interactions. Consequently, the origin of spin-glass behaviour in Y{$_2$}Mo{$_2$}O{$_7$}---in which magnetic Mo$^{4+}$ ions occupy a frustrated pyrochlore lattice with minimal compositional disorder---has been a longstanding question. Here, we use neutron and X-ray pair-distribution function (PDF) analysis to develop a disorder model that resolves apparent incompatibilities between previously-reported PDF, EXAFS and NMR studies and provides a new and physical mechanism for spin-glass formation. We show that Mo$^{4+}$ ions displace according to a local ``2-in/2-out" rule on each Mo$_4$ tetrahedron, driven by orbital dimerisation of Jahn-Teller active Mo$^{4+}$ ions. Long-range orbital order is prevented by the macroscopic degeneracy of dimer coverings permitted by the pyrochlore lattice. Cooperative O$^{2-}$ displacements yield a distribution of Mo--O--Mo angles, which in turn introduces disorder into magnetic interactions. Our study demonstrates experimentally how frustration of atomic displacements can assume the role of compositional disorder in driving a spin-glass transition.
\end{abstract}


\maketitle

In a spin-glass transition, spins freeze into a metastable arrangement without long-range order \cite{Blundell_2001}. It is generally accepted that two conditions must be satisfied for a spin-glass transition to occur: interactions between spins must be disordered and these interactions must also be frustrated \cite{Binder_1986,Huang_1985}. In canonical spin glasses---\emph{e.g.}, dilute magnetic alloys such as Cu$_{1-x}$Mn${_x}$  \cite{Lundgren_1983} and site-disordered crystals such as Fe$_2$TiO$_5$ \cite{Atzmony_1979}---the nature of structural disorder and its coupling to magnetism is well understood. However, spin-glass behaviour is also observed in well-ordered crystals where the geometry of the magnetic lattice alone can generate frustration (see, \emph{e.g.}, \cite{Schiffer_1995,Wiebe_2003, Benbow_2009}). Here, the mechanism of spin-glass formation poses an important challenge for theory \cite{Saunders_2007,LaForge_2013}. The prototypical material that shows this anomalous behaviour is Y$_{2}$Mo$_2$O$_7$---a system with apparently unremarkable levels of structural disorder, but with thermodynamic properties indistinguishable from canonical spin glasses (for a review, see \cite{Gardner_2010}).

The structure and dynamics of Y$_{2}$Mo$_2$O$_7$ have been extensively studied. The conventional nature of the spin-glass transition (freezing temperature $T_\mathrm{f}=22$\,K \cite{Gingras_1997}) has been shown by exhaustive thermodynamic measurements, including non-linear susceptibility \cite{Gingras_1997,Gingras_1996}, specific heat \cite{Blacklock_1980,Raju_1992}, a.c.\ susceptibility \cite{Miyoshi_2000}, and thermo-remanent magnetization \cite{Ali_1992,Dupuis_2002,Ladieu_2004}. Inelastic neutron scattering \cite{Gardner_1999}, muon-spin rotation ($\mu$SR) \cite{Dunsiger_1996}, and neutron spin-echo studies \cite{Gardner_2001,Gardner_2004} reveal a reduction in the spin-relaxation rate as $T_\mathrm{f}$ is traversed. 
Neutron diffraction measurements show that the average structure is well-described by the ordered pyrochlore model (space group $Fd\bar{3}m$) both below and above $T_\mathrm{f}$ \cite{Reimers_1988,Greedan_2009}. In this structure, the average positions of magnetic Mo$^{4+}$ ions describe a network of corner-sharing tetrahedra [Fig.~1(a)]. There are two inequivalent O sites, O1 and O2; each Mo$^{4+}$ is octahedrally coordinated by O1 [Fig.~1(b)], and each Y$^{3+}$ is coordinated by six O1 and two O2 [Fig.~1(c)]. Each pair of Mo neighbours is bridged by a single O1, forming the main magnetic superexchange pathway. The degree of site-mixing and O non-stoichiometry is too small to be measured \cite{Reimers_1988,Greedan_2009} and is calculated to be minimal ($\sim$1\%) \cite{Minervini_2002}.

\begin{figure}
\begin{center}
\includegraphics{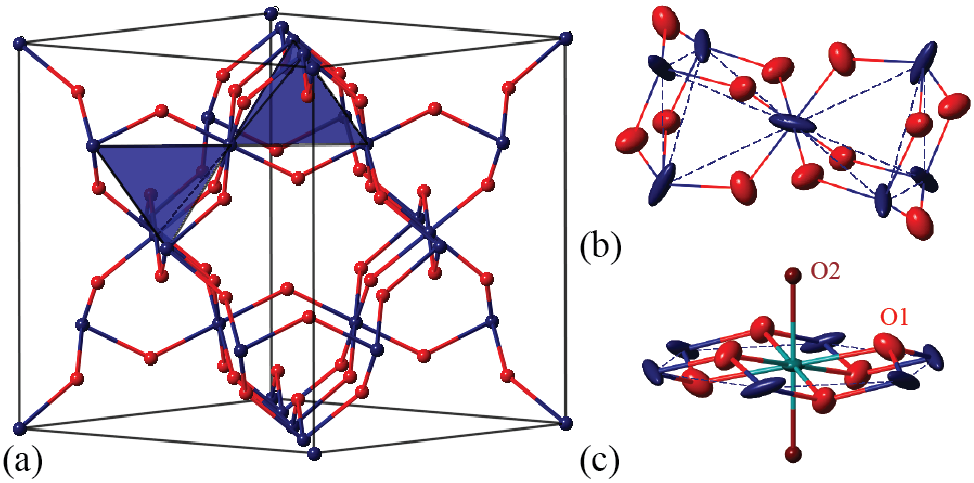}
\end{center}
\caption{\label{fig1} (a) Crystal structure of Y{$_2$}Mo{$_2$}O{$_7$} generated from the model of Ref.~\onlinecite{Greedan_2009}: Mo in dark blue, O in red, and Y omitted for clarity. (b) Coordination environment of Mo$^{4+}$ shown with displacement ellipsoids at 50\% probability. Prolate Mo$^{4+}$ displacement ellipsoids point along the local-$\langle 111 \rangle$ axes (towards the centres of adjacent tetrahedra). (c) Coordination environment of Y$^{3+}$ (green).}
\end{figure}

A plausible mechanism for the spin-glass transition invokes local Mo$^{4+}$ displacements to generate variation in magnetic interactions \cite{Keren_2001}. Arguably the clearest experimental signature of anomalies on the Mo site is its large and anisotropic atomic displacement parameter (ADP) obtained from Rietveld refinement to powder neutron diffraction data \cite{Greedan_2009} [Fig.~1(b)]. Consistent with this observation, an EXAFS study (Mo and Y $K$-edges) showed a large static variance in Mo--Mo distances, $\sigma_\mathrm{stat}^2=0.026(5)$\,\AA$^2$ \cite{Booth_2000}. Two \textsuperscript{89}Y NMR studies revealed a distribution of Y environments that was interpreted in terms of a local distortion of the Mo site \cite{Keren_2001,Ofer_2010}, an interpretation supported by $\mu$SR experiments \cite{Sagi_2005}.
There is a problem with this proposal, however: a state-of-the-art study using neutron pair-distribution function (PDF) analysis found no evidence for a local splitting of the Mo site \cite{Greedan_2009}. Instead, it showed pronounced variation in Y--O1 distances consistent with a local splitting of the O1 site---a result interpreted as directly contradicting the EXAFS study \cite{Greedan_2009,Booth_2000}.

In this Letter, we argue that a disorder model that explains these apparently contradictory results holds the key to understanding the mechanism of spin-glass formation in Y$_{2}$Mo$_2$O$_7$.
We critically assess the validity of the key assumption \cite{Greedan_2009} that the absence of obvious Mo--Mo splitting in the PDF implies that Mo off-centring does not occur. We show that the PDF is actually better represented by a structural model in which Mo does off-centre and this off-centring is coupled to O1 displacements. The Mo displacements we obtain are consistent with EXAFS \cite{Booth_2000} and are sufficiently large to give the requisite variation in magnetic interactions. We show that Mo displacements obey a ``2-in/2-out" rule on each Mo$_4$ tetrahedron that can be interpreted as the formation of orbital dimers. The magnetic interaction within a dimer is different to the interactions between dimers, and the pyrochlore lattice permits a macroscopic degeneracy of disordered dimer coverings. Our model therefore explains the disorder in magnetic interactions that may drive the spin-glass transition.

Our paper is structured as follows. After describing our experimental procedures, we first show that we can reproduce the best fit to PDF data obtained previously \cite{Greedan_2009}. We then describe the simplest orbital-dimer model, which can be interpreted as a crystalline approximant to a more complex disordered state. 
We demonstrate that this new local-structure model yields a better fit to PDF data than the previous model \cite{Greedan_2009}, for the same number of structural parameters. Finally, we show that our model is consistent with previous experimental studies \cite{Booth_2000,Keren_2001,Ofer_2010} and with theoretical requirements for spin-glass formation.

A polycrystalline sample of Y{$_2$}Mo{$_2$}O{$_7$} (8\,g) was prepared by firing stoichiometric amounts of Y{$_2$}O{$_3$} and MoO{$_2$} for 12 hours at 1400\,$^{\circ}$C using CO/CO{$_2$} as buffer gas \cite{Sato_1986}. The value of $T_\mathrm{f}$ determined from measurements of the field-cooled and zero-field-cooled magnetization was consistent with the previously-reported value (22\,K) \cite{Greedan_1986}. Neutron total-scattering data were collected on the recently-upgraded POLARIS instrument at ISIS \cite{Hull_1992} and X-ray scattering data were collected on the 11-ID-B beamline at the Advanced Photon Source. The neutron data were normalized using a vanadium standard and corrected for background scattering and absorption using the GUDRUN program \cite{Soper_2011}. The reciprocal-space range used was $1.0\leq Q\leq25\,\mathrm{\AA}^{-1}$ for the X-ray data and  $0.7\leq Q\leq45\,\mathrm{\AA}^{-1}$ for the neutron data. This excludes most of the diffuse magnetic contribution to the neutron data, which peaks at $0.4\,\mathrm{\AA}^{-1}$ \cite{Gardner_1999} and was barely observed above background in our high-$Q$-optimised measurement. All PDF analysis was carried out using the PDFGui program \cite{Farrow_2007} by refinement to both X-ray and neutron PDFs over the range $0 < r < 10$\,\AA, with a weighting factor chosen so that both datasets sets contributed approximately equally to the refinement. 

\begin{figure}
\begin{center}
\includegraphics{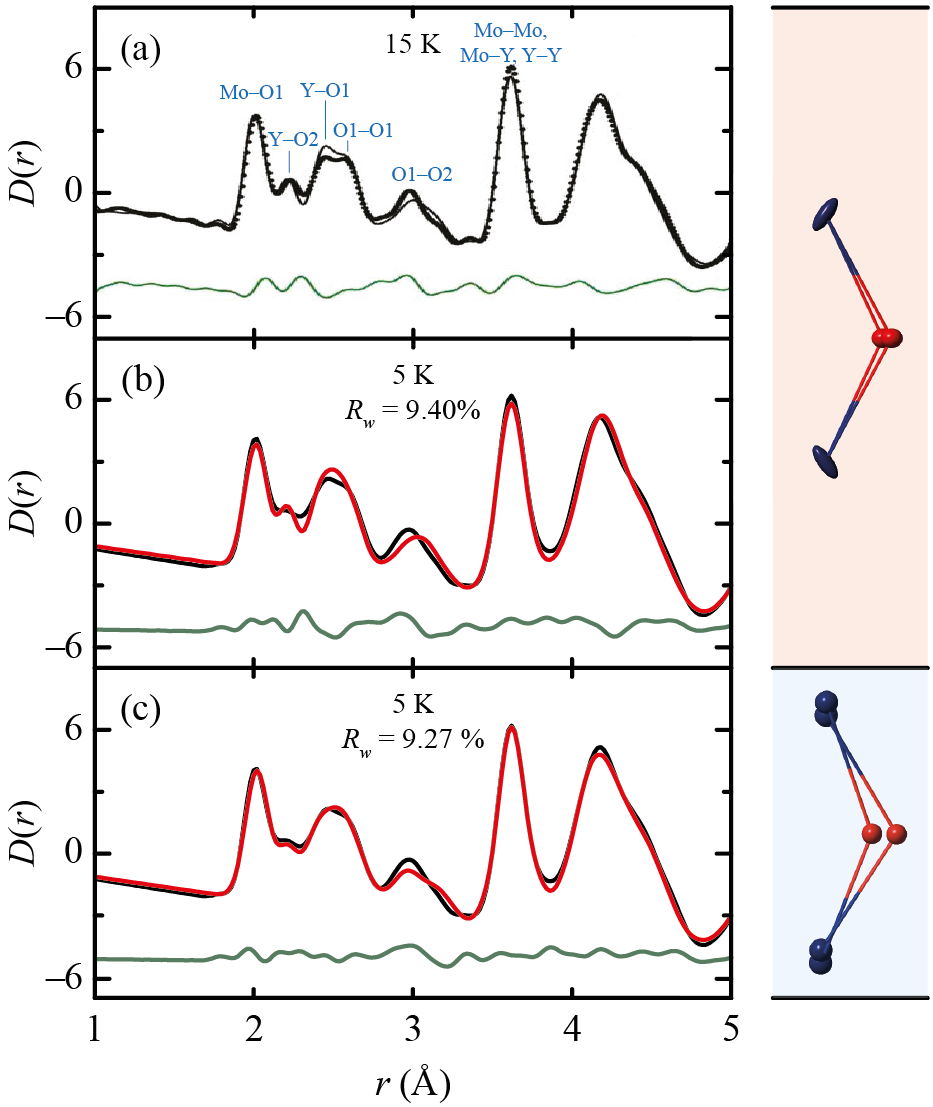}
\end{center}
\caption{\label{fig2} (a) Reproduction of the fit to neutron PDF data shown in Fig.~8 of Ref.~\citenum{Greedan_2009} for the split-site model described in the text. (b) Fit of the split-site model to our neutron PDF data. (c) Fit of the orbital dimer model ($I4_{1}md$; described in the text) to our neutron PDF data. In (a), data are shown as points and fit as a line; in (b) and (c), data are shown as black lines, fits as red lines, and data$-$fit as green lines. The extrema of Mo--O1--Mo conformations accounted for by each model are shown beside the corresponding fit.}
\end{figure}

In the PDF study of Ref.~\onlinecite{Greedan_2009}, the local structure of Y$_2$Mo$_2$O$_7$ is modelled using the conventional $Fd\bar3m$ (average) structure with the O1 site partitioned equally into two sites (O1a,O1b) at fractional coordinates $(x,1/8,1/8)$; we refer to this as the ``split-site model". Our starting point was to reproduce the quality of PDF fit obtained in \cite{Greedan_2009} using this model and our newly-collected X-ray + neutron data sets. We found that stable refinement required the use of constraints on the anisotropic displacement parameters for the O1a/O1b sites---hence the number of parameters in our refinement is marginally smaller than in \cite{Greedan_2009} (see SI for further details). Nevertheless the quality of our fit for this split-site model and the form of the difference function (data $-$ fit) are both essentially indistinguishable from that reported in Ref.~\onlinecite{Greedan_2009} [Fig.~\ref{fig2}(a,b)]; a comparison of refined parameter values is given in Table~\ref{table1}. We conclude that our sample, data, and refinement procedures are entirely consistent with \cite{Greedan_2009}.

 \begin{table}
\begin{center}
\caption {Structural parameters for our PDF refinements using the split-site and orbital dimer models described in the text. Reference values from the split-site refinements of Ref.~\onlinecite{Greedan_2009} are shown in italics (* = Rietveld and $\dagger$ = PDF refinements; note that the level of consistency in cell parameters is typical for Rietveld/PDF comparisons). Lattice parameters are given in \AA\ and displacement parameters in 10$^{-2}$\,\AA$^{2}$.}
\begin{tabular}{lr@{.}lr@{.}l | lr@{.}l}
\hline\hline
 \multicolumn{5}{c|}{\textbf{Split-site model \cite{Greedan_2009}}} & \multicolumn{3}{c}{\textbf{Orbital dimer model}}\\
 \hline
$a$ & 10&2255(10)  & \emph{10}&\emph{20753(2)*} & $c_\mathrm{t}\equiv\sqrt{2}a_{\mathrm t}$ & 10&2269(10)\\
$x$(O1a) & 0&3320(6) &\emph{0}&\emph{3305(12)$\dagger$} & $y$(Mo) & 0&7371(8)\\
$x$(O1b) & 0&3446(6) & \emph{0}&\emph{3465(13)$\dagger$} & $z$(O1a) & 0&6938(6)\\
\textit{U}$_{11}$(Y) & 0&512(15) & \emph{0}&\emph{471(7)*} & $x$(O1b) & 0&7184(5)\\
 \textit{U}$_{12}$(Y) & $-$0&129(19) & \emph{$-$0}&\emph{136(9)*} & $y$(O1b) & 0&7910(5)\\
 \textit{U}$_{11}$(Mo) & 1&39(5) & \emph{1}&\emph{1(1)*} & $z$(O1b) & 0&2467(4)\\
 \textit{U}$_{12}$(Mo) & 0&93(5) & \emph{0}&\emph{83(2)*} & $z$(O1c) & 0&2709(6) \\
 \textit{U}$_{11}$(O1)& 1&02(11) & \emph{1}&\emph{45(2)*} & \textit{U}$_\mathrm{iso}$(Y) & 0&355(11)\\
 \textit{U}$_{22}$(O1)& 0&81(3) & \emph{0}&\emph{66(1)*} & \textit{U}$_\mathrm{iso}$(Mo)& 1&14(5)\\
\textit{U}$_\mathrm{iso}$(O2) & 0&55(3) & \emph{0}&\emph{32(2)*} & \textit{U}$_\mathrm{iso}$(O)& 0&89(2)\\
 \hline\hline
			\end{tabular}
		\label{table1}
	\end{center}
\end{table}  


We now propose an alternative model of static disorder in Y$_{2}$Mo$_2$O$_7$. Our starting-point is the observation that Mo ADPs are strongly elongated along the local-$\langle 111 \rangle$ axes in the average-structure model [Fig.~1(b)]. This result suggests that Mo$^{4+}$ ions are locally displaced towards or away from the centres of Mo$_4$ tetrahedra. Physically-reasonable mechanisms for Mo displacements, such as charge polarisation and orbital interactions \cite{Shinaoka_2013}, require that displacements are not random but coupled. There are two basic possibilities for this coupling. It may require that the Mo displacements on a tetrahedron either all point towards its centre or all point away (``4-in/4-out" state); alternatively, it may require that two Mo displacements point towards the centre and two point away (``2-in/2-out" state). The first pattern of displacements is necessarily ordered on the pyrochlore lattice, and is ruled out in Y$_{2}$Mo$_2$O$_7$ by the absence of global symmetry lowering. By contrast, the second pattern need not be ordered at all. In fact, if all 2-in/2-out states were equally probable, there would be a macroscopic degeneracy of Mo-displaced configurations analogous to the degeneracy of proton configurations in cubic water ice \cite{Pauling_1935}---\emph{i.e.}, an orbital-ice state. 

To develop a model that can be directly compared with the split-site model in terms of the ability to account for the experimental PDF, we consider an ordered approximant to the ensemble of possible disordered 2-in/2-out states. For distances within a single unit cell an ordered 2-in/2-out state will show essentially the same PDF as the disordered states, because both satisfy the same local 2-in/2-out constraint. The highest-symmetry subgroup of $Fd\bar{3}m$ that permits 2-in/2-out Mo displacements is $I4_{1}md$. This structure is shown in Fig.~3(a), and relates to $Fd\bar{3}m$ in the same way as the proton-ordered ice phase XIc relates to proton-disordered cubic ice Ic \cite{Raza_2011}. The tetragonal unit cell has dimensions $a_{\mathrm{t}}=a/\sqrt{2}$ and $c_{\mathrm{t}}=a$. Both Mo and Y occupy the $8b$ site with coordinates $(0,y,z)$, O2 occupies the $4a$ site with coordinates $(0,0,z)$, and the O1 site is split into three sites: O1a and O1c also occupy the 4$a$ site and O1b occupies the $16c$ site with coordinates $(x,y,z)$. The point symmetry of the Mo site is lowered from ${\bar3}m$ to $m$, which is consistent with a Jahn-Teller-type distortion of the $d^2$ electronic configuration of Mo$^{4+}$.

We reduce the number of refined parameters in our $I4_{1}md$ model to match that in the split-site model in the following way. First, we constrain the ratio of unit-cell dimensions $c_{\mathrm{t}}=\sqrt{2}a_{\mathrm{t}}$, so that the parent cell remains metrically cubic. Second, the small displacement parameters of the Y and O2 sites indicate that they are not strongly disordered [Fig.~1(c)]; we therefore fix the Y and O2 positions to their average values of (0,$\frac{3}{4}$,$\frac{1}{8}$) and (0,0,0), respectively. Finally, the $I4_{1}md$ structure now allows for Mo displacements within the plane that contains the cubic local-$\langle 111 \rangle$ axis (towards or away from the tetrahedron centre) and local-$\langle 110 \rangle$ axis (towards or away from neighbouring Mo atoms lying on the same mirror plane). Consequently, two parameters are needed to define the Mo displacement direction, which would increase the number of parameters beyond the split-site model. Such a refinement was stable, however, and showed that the dominant displacement direction is in fact the local-$\langle 110 \rangle$. Subsequently, the local-$\langle 111 \rangle$ component was fixed at zero ($z({\mathrm{Mo}})\equiv \frac{5}{8}$), so that the number of refined structural parameters was identical to the split-site model. We will refer to this constrained $I4_1md$ model as the ``orbital dimer'' model.

We refined this model against neutron and X-ray PDF data simultaneously using PDFGui \cite{Farrow_2007}. The fit to neutron data is presented in Fig.~\ref{fig2}(c); values of refined parameters are given in Table~\ref{table1}, and the fit to X-ray data is given in SI. The quality of fit for the orbital dimer model is higher than for the split-site model ($R_{\mathrm{wp}}=9.27\%$ \emph{vs} $9.40\%$) despite employing the same number of structural parameters. Moreover, the $I4_{1}md$ model obviously accounts more convincingly for the low-$r$ peaks. Further evidence for the improved model quality comes from the refinement statistics: whereas the split-site model showed strong covariance ($>80\%$) between the $x$(O1a), $x$(O1b), and $U$$_{11}$(O1) parameters, the orbital dimer refinement resulted in no anomalous covariance terms. In fact, a robust refinement could also be achieved with a larger number of structural parameters (see SI). We therefore conclude that the absence of visible splitting of the Mo--Mo peak cannot be said to be inconsistent with Mo off-centring. On the contrary, our model includes Mo displacements and robustly yields an improved fit to experimental data [Fig.~2c].

\begin{figure}
\begin{center}
\includegraphics{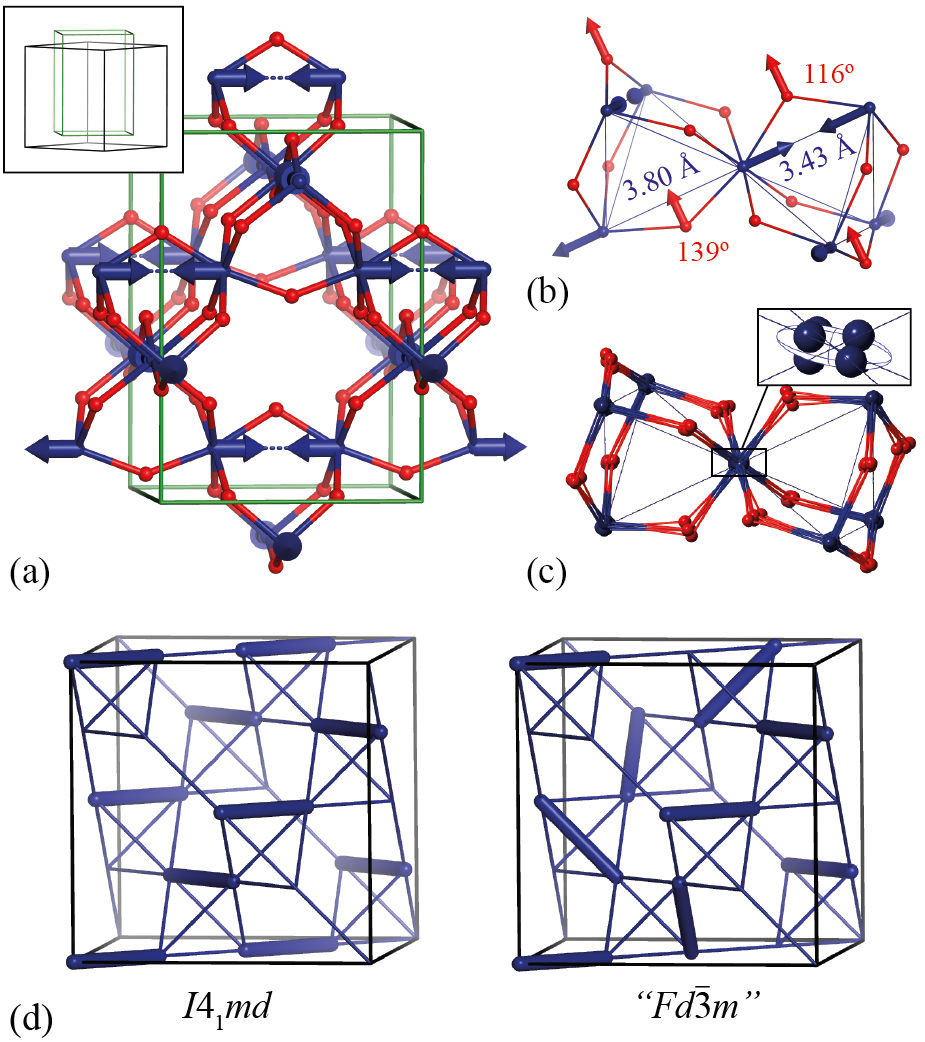}
\end{center}
\caption{\label{fig3} (a) Representation of the orbital dimer model obtained from PDF refinements: O1 atoms in red, Mo displacements are shown as blue arrows, and Mo--Mo dimers as blue dashed lines. The inset shows the relationship between $I4_1md$ and $Fd\bar{3}m$ cells. (b) Local Mo environment in this model, with Mo displacement directions shown as blue arrows and dominant O1 displacement directions shown as red arrows. (c) Distribution of atomic positions obtained by imposing the full symmetry of the average structure. The superposition of displaced Mo sites is compared in the inset to the prolate displacement ellipsoid obtained from Rietveld refinement. (d) Comparison of ordered and disordered coverings of the pyrochlore lattice by Mo--Mo dimers (thick blue lines). The left panel shows the ordered dimer covering for the $I4_1md$ structure, which breaks cubic symmetry. The right panel shows an example of a disordered dimer covering that more closely approximates the true orbital ice state; spatial averaging over such dimer coverings preserves cubic symmetry.}
\end{figure}

The physical interpretation of the orbital dimer model is represented in Fig.~\ref{fig3}. In every Mo$_4$ tetrahedron, two Mo are displaced towards each other and two away from each other; in each case, the displacement magnitude is $0.093(6)\,\mathrm{\AA}$ and the displacement direction is along the line connecting the Mo pair [Fig.~\ref{fig3}(a)]. The Mo displacements thus describe a covering of the pyrochlore lattice by Mo--Mo dimers. The O1 displacements are correlated with Mo displacements as shown in Fig.~\ref{fig3}(b). When neighbouring Mo displace towards each other, the bridging O1 displaces away from the dimer and the Mo--O1--Mo angle decreases; conversely, when neighbouring Mo displace away from each other the Mo--O1--Mo angle increases. The effect of O1 displacements is to keep all Mo--O1 distances essentially the same---a result that is perhaps unsurprising on electrostatic grounds but nevertheless explains the lack of visible Mo--O1 splitting in the PDF. Fig.~\ref{fig3}(c) shows the distribution of atomic positions obtained by superposing the full $Fd\bar3m$ symmetry of the average structure on the local structure model refined against PDF data. The distribution of local-$\langle 110 \rangle$ Mo displacements has a larger component parallel to the local-$\langle 111 \rangle$ axis than perpendicular to it, consistent with the prolate Mo displacement ellipsoid in the average-structure model [Fig.~\ref{fig1}(b) and inset to Fig.~\ref{fig3}(c)]; details are given in SI. Finally, we stress that the ordered $I4_1md$ dimer covering [left panel of Fig.~\ref{fig3}(d)] is used as an \emph{approximant} to disordered dimer coverings that show the same local displacement patterns. There is a macroscopic degeneracy of disordered dimer coverings that reproduce the observed $Fd\bar{3}m$ symmetry on spatial averaging; the right panel of Fig.~3(d) shows a disordered example that would more closely approximate the true structure. 
This is the same degeneracy responsible for the unusual physics of static and dynamic spin-ice materials \cite{Bramwell_2001,Zhou_2008}, cubic water ice \cite{Pauling_1935}, and charge-ice materials \cite{Shoemaker_2010,Fairbank_2012}.

Crucially, the orbital model is consistent with previous experimental studies and theoretical requirements for spin-glass formation. The O1 displacement gives rise to 30 distinct Y environments, consistent with the broad resonance observed in \textsuperscript{89}Y NMR studies \cite{Keren_2001,Ofer_2010}. 
The Mo displacements split the average Mo--Mo distance into long, short, and intermediate distances in the ratio 1:1:4. 
The variance of Mo--Mo distances is 0.012(2)\,\AA${^2}$, in qualitative agreement with the EXAFS result (0.026(5)\,\AA${^2}$); moreover, the displacement direction (parallel to Mo--Mo pairs) is consistent with EXAFS \cite{Booth_2000}. This agreement is encouraging because EXAFS should be more sensitive to Mo displacements than the PDF, in which Mo--Mo, Y--Y and Mo--Y peaks overlap and our intermediate Mo--Mo distance is similar to the average Mo--Mo distance.
It has been shown that small ($\sim$10\%) variations in exchange interactions may induce spin-glass transitions in geometrically-frustrated magnets \cite{Saunders_2007}. In Y$_{2}$Mo$_2$O$_7$, variations in bond angle of $9^{\circ}$ are calculated to yield factor-of-two variations in the strength of the magnetic coupling \cite{Greedan_2009}. Our results show that the Mo--O1--Mo angle actually varies by a significantly larger amount, 23$^{\circ}$ [Fig.~3(b)]. Moreover, the arrangement of Mo--O1--Mo angles---and therefore magnetic couplings---is disordered, because the dimer covering is disordered [Fig.~3(d)]. Hence, Mo dimerisation presents a clear mechanism for spin-glass freezing. 
Theoretical studies suggest that Mo dimerisation may be explained by an orbital (Jahn-Teller) mechanism. A recent study parameterised a three-orbital Hubbard model using DFT \cite{Shinaoka_2013}, and concluded that spin and orbital degrees of freedom are strongly coupled for the $d^2$ configuration and many energetically-similar spin-orbital excited states exist if the ground state is 2-in/2-out. A second DFT study relaxed the orbital and lattice configuration for different spin structures and obtained different lattice distortions in each case \cite{Silverstein_2014}. These studies provide compelling evidence for a strong coupling of spin, orbital, and lattice degrees of freedom. We therefore interpret the 2-in/2-out pattern of Mo displacements as the formation of orbital dimers, driven by the Jahn-Teller activity of the $d^{2}$ electronic configuration. 
 
Our results identify Y$_2$Mo{$_2$}O{$_7$} as a rare example of a spin glass that is not driven by random compositional or site disorder \cite{Goremychkin_2008}.
Instead, strong orbital interactions drive the formation of Mo--Mo dimers, which do not order because the frustrated topology of the pyrochlore lattice supports disordered dimer coverings. While spin-glass formation due to non-random interactions has been studied theoretically \cite{Villain_1977}, to the best of our knowledge this represents the first such example for a real spin glass. Yet, dimer states play a key role in determining other properties of materials. In LiV{$_2$}O{$_4$}, charge dimerisation on V$_4$ tetrahedra of the pyrochlore network may explain the observed heavy-fermion behaviour \cite{Fulde_2001}. In Ba{$_2$}YMoO{$_6$}, Mo$^{5+}$ spins freeze into a disordered arrangement of spin-singlet dimers \cite{Vries_2010}, while in CuIr{$_2$}O{$_4$} Ir$^{3+}$/Ir$^{4+}$ charge ordering occurs simultaneously with spin dimerisation \cite{Radaelli_2002}. We anticipate that a dimer model could be applicable to other molybdate pyrochlores; \emph{e.g.}, Lu$_2$Mo$_2$O$_7$ \cite{Clark_2014} and Tb$_2$Mo$_2$O$_7$ \cite{Gaulin_1992,Singh_2008}. 
Our results suggest three promising directions for future work on Y$_2$Mo{$_2$}O{$_7$}. First, a comprehensive single-crystal diffuse scattering study would indicate the extent to which a particular subset of dimer coverings is preferentially selected. Second, analysis of the distribution of magnetic couplings may allow a consistent interpretation of puzzling magnetic diffuse-scattering data \cite{Silverstein_2014}. Third, the orbital-dimer state we propose should collapse at sufficiently high temperature; \emph{i.e.}, an orbital-ice to orbital-liquid transition may be anticipated. 

We thank A. Simonov, J.~R. Stewart, M. Mourigal, C.~R. Wiebe, H.~J. Silverstein, M.~J~.P. Gingras, F. Flicker, and J.~S. Gardner for useful discussions. We acknowledge the Rutherford Appleton Laboratory for access to the ISIS Neutron Source. This research used resources of the Advanced Photon Source, a U.S. Department of Energy (DOE) Office of Science User Facility operated for the DOE Office of Science by Argonne National Laboratory under Contract No. DE-AC02-06CH11357. P.M.M.T., J.A.M.P., and A.L.G. acknowledge financial support from the STFC, E.P.S.R.C.\ (EP/G004528/2), and the E.R.C. (Grant Ref: 279705).



\end{document}